\documentstyle[12pt,twoside,fleqn,espcrc1,psfig]{article}

\newcommand{ \be }{\begin{equation}}
\newcommand{ \ee }{\end{equation}}

\newcommand{ \bea }{\begin{eqnarray}}
\newcommand{ \eea }{\end{eqnarray}}

\newcommand{ \la }{\langle}
\newcommand{ \ra }{\rangle}

\newcommand{\AmS}{{\protect\the\textfont2
  A\kern-.1667em\lower.5ex\hbox{M}\kern-.125emS}}

\title{Anisotropic flow of identified particles in Au+Au collisions 
at AGS energy}

\author{Sergei A. Voloshin\address{Physikalisches Institut, 
University of Heidelberg, Heidelberg, Germany}\thanks{On leave from 
Moscow Engineering Physics Institute, Moscow, 115409,  Russia}
 for the E877 Collaboration\address{
BNL, GSI, INEL, University of Heidelberg, McGill University, 
Pittsburgh University, 
SUNY Stony Brook, University of S\~ao Paulo, Wayne State University}
}

\begin{document}
\maketitle

\begin{abstract}
Anisotropic flow of protons, $\pi^{\pm}$, $K^+$, deuterons, tritons,  
$^3$He, and $^4$He is analyzed as a function of transverse momentum for
different centralities of the collision.
\end{abstract}

\section{E877 experimental setup and flow analysis}

The E877 experimental setup is discussed in detail 
elsewhere~\cite{l877flow1,l877flow2,l877flow3,lwessels}.
In the current analysis, for the reaction plane determination
and the evaluation of the reaction plane resolution, we exploit
the almost 4$\pi$ calorimeter coverage of the apparatus 
(the Target Calorimeter (TCal)
covers the pseudorapidity range $-0.5<\eta<0.8$, the Participant Calorimeter 
(PCal) $0.8<\eta<4.2$).
Charged particles, emitted in the forward direction 
($ -134 < \theta_{horizontal} < 16 $ mrad, $ -11 <
\theta_{vertical} < 11 $ mrad), are analyzed by a high resolution
magnetic spectrometer. 
The average momentum
resolution is $\Delta p/p \approx$3\% limited by multiple scattering.
A time-of-flight hodoscope located behind the tracking chambers
(10~m from the target)
provides time-of-flight information with a typical resolution of 85~ps.
For the identification of charge=2 particles we also use the Forward
Scintillator array located approximately 33~m downstream of the target.

The centrality of the collisions is determined by the transverse energy
deposited in the PCal (see Table 1).

\medskip
\parbox{8cm}{
\begin{tabular}{|c||c|c|c|c|}
 \hline 
 Centrality & PCal $E_t$ (GeV)& $\sigma_{top}/\sigma_{geo}$
&  $b$ (fm)
 \\ 
\hline
1 & 150 -- 200 & 0.23 -- 0.13 & 5 -- 7\\
2 & 200 -- 230 & 0.13 -- 0.09 & 4 -- 5\\
3 & 130 -- 270 & 0.09 -- 0.04 & 3 -- 4\\
4 &  $>$ 270   &$<$ 0.04       &  $<$ 3 \\
\hline
\end{tabular}
}
\hspace{2cm}\parbox{5cm}{
{Table 1. Centrality regions. The impact parameter range was estimated 
in accordance with $\sigma_{top}/\sigma_{geo} \approx(b/2R)^2$.} 
}

\medskip
To describe the anisotropy in particle production we use a Fourier expansion
of the azimuthal distributions~\cite{lvzh,lolli} (see 
also~\cite{l877flow2,l877flow3}), where the anisotropic flow signal 
is represented by $v_n$, the amplitudes of different harmonics.
In this analysis we discuss mostly directed flow, $v_1(p_t)$, with emphasis on
the dependence on the particle transverse momentum.
All results shown below are corrected for the reaction plane resolution 
in accordance with the procedure described in~\cite{l877flow2,l877flow3,lolli}.

\section{Results}

{\it Protons.} 
The directed flow signal, $v_1(p_t)$, of protons almost linearly depends on
$p_t$ (see Figures~1 and~4; for a complete set of proton
and pion data and also results on $v_2$ see~\cite{l877flow3}).
This behavior naturally follows from the assumption that directed flow
is a consequence of the sideward motion of the thermal source with
velocity $\beta_x$~\cite{l877flow3}).
Small deviations from a linear dependence at low $p_t$ can be explained
in this case by taking into account the transverse radial expansion of 
the source~\cite{lvrd}.
Note that in such a picture the shape of $v_1(p_t)$ depends
on the magnitude of radial expansion velocity $\beta_r$, providing
the possibility to study radial expansion by means of an anisotropic flow
analysis.
Using this model we fit our proton data under the assumption that
the source temperature is fixed, $T=110$~MeV.
The results of the fit are shown in Fig.~1 and the fit 
parameters (directed and radial velocities, $\beta_x$ and $\beta_r$, in
this case) are presented in Fig.~2.
 

\bigskip
\psfig{figure=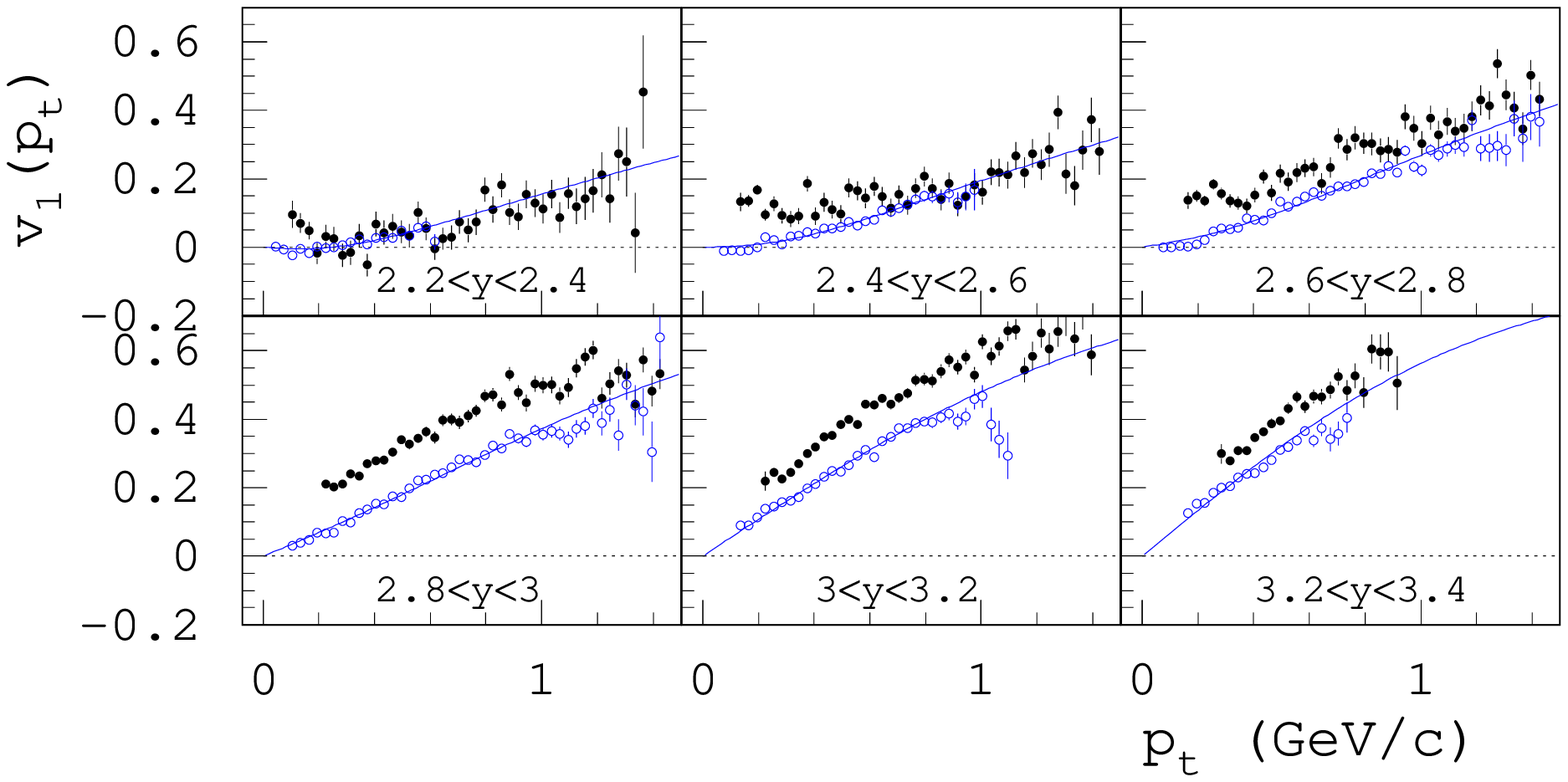,height=8.cm}
\vspace{-13mm}
{\small Figure 1.  Directed flow of protons (open symbols)
and deuterons (solid symbols) for collisions in  centrality region 2. 
A (solid line) fit to the proton data was performed 
taking into account the radial expansion of the source.}
\smallskip

\vspace{-8mm}
\psfig{figure=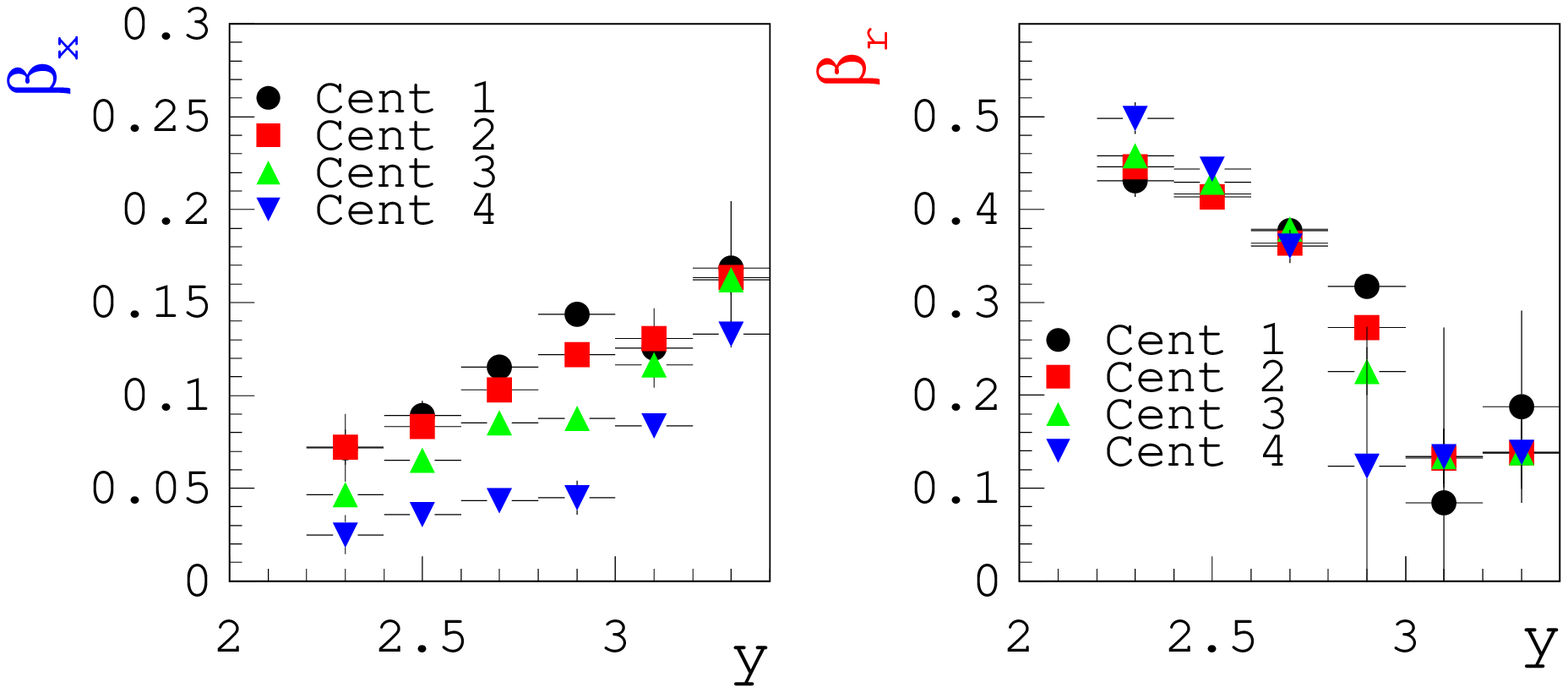,height=7.5cm}
\vspace{-12mm}
\centerline{{\small Figure 2. Directed and radial velocities of the
proton source from the fit to $v_1(p_t)$.}}
\smallskip

{\it Charged pions}.
Directed flow of $\pi^+$ and $\pi^-$ (see Figures 6 and 7
in~\cite{l877flow3}) is very similar for $p_t>$100~MeV/c (being negative
at $100<p_t<300$--500~MeV/c with values $v_1\approx -0.05$ -- $-0.1$, and
becoming positive at higher $p_t$ values), and is different at 
$p_t<$100~MeV/c (at $p_t<40$~MeV/c negative pions exhibit positive flow!).
Such behavior at least qualitatively can be explained by 
four factors: 1) shadowing by the comoving nucleons, 
2)~transverse motion of the source (the same as for protons),
3) Coulomb interactions with comoving protons, and
4) $\Delta$ (and, to some extent, $\Lambda$) decays.
Note that the Coulomb effect in this case is different from the one
widely discussed in the literature, namely, from the distortion 
of the spectra due to the central potential.
In our case the comoving protons (mostly spectators) are shifted with respect
to the pions in the transverse plane;  then Coulomb interaction
would result in directed flow of negative pions in the direction of proton
flow, and of positive pions in the opposite direction.  
A very rough
estimate of the signal in the case of a static Coulomb field result in
$v_1(p_t \rightarrow 0) \approx \la r\ra/a$, where $\la r \ra$ is the
relative source shift and $a$ is the (pion)  Bohr radius.  Note that
the signal in this case is proportional to the particle mass (inversely
proportional to Bohr radius); then ``Coulomb flow'' of kaons should be even
larger than that of pions.  

Note also the rather non-trivial
dependence on $p_t$ of the flow signal $v_1$ of pions from decays of $\Delta$
resonances. Simple considerations show that the ``kinematics'' of such
kind of flow under assumption that $\Delta$ flow follows the proton flow
(moving thermal source) is exactly the same as in the picture of
``a transversely moving and radially expanding thermal source''~\cite{lvrd}.  
It means that pions from $\Delta$ decays would exhibit negative flow at
low $p_t$ and follow protons at high transverse momentum.

{\it Kaons}.
The magnetic field polarity used for the analyzed data set 
provided good acceptance only for positively charged kaons.
The observed flow signal (Fig.~3) is very small (compatible with zero)
for $p_t>200$~MeV/c. 
In the low $p_t$ region strong (negative) flow is observed.
It could be due to Coulomb interaction with comoving protons 
(see above). 
The currently undergoing analysis of the (different field polarity
data) $K^-$ flow will clarify the role of Coulomb interaction in kaon flow. 

\vspace{-2mm}
\centerline{
\psfig{figure=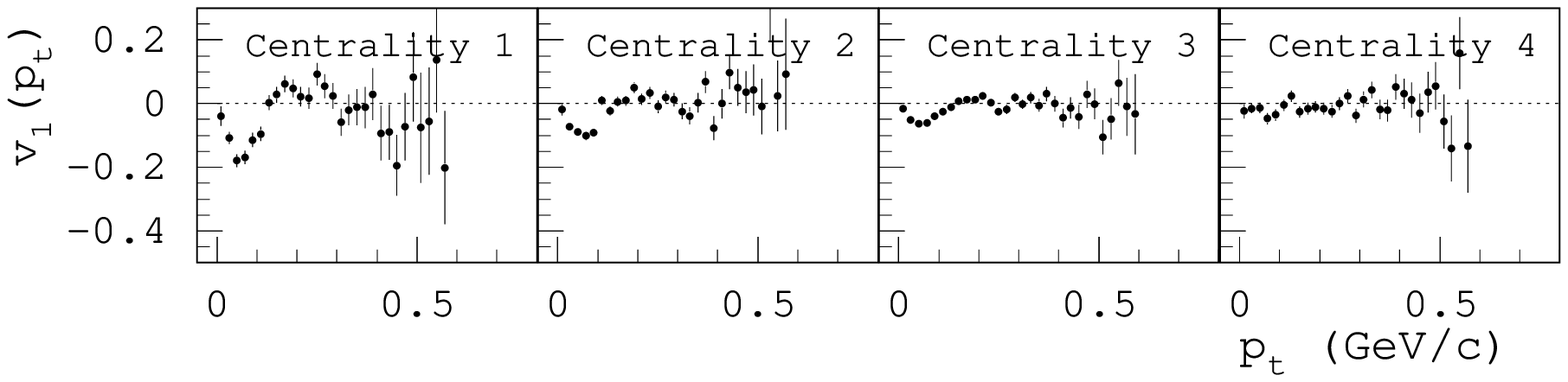,height=57mm}
}
\vspace{-1mm}
\centerline{{\small Figure 3.  $K^+$ directed flow in the rapidity
region $2.4<y<2.8$}}
\smallskip

{\it Deuterons}.
Deuterons exhibit strong directed flow (see Figures~1 and 4).
Note that in the {\em moving thermal source}
model~\cite{l877flow3,lvrd} $v_1(p_t)$
in first order does not depend on the mass of the particle; 
then $v_1^{deuteron}(p_t)=v_1^{proton}(p_t)$.
Also, in a simple {\em coalescence} model (without volume effects) 
$v_1^{deuteron}(p_t) \approx 2v_1^{proton}(p_t/2)\approx v_1^{proton}(p_t)$.
The observed significant excess of deuteron flow in comparison with
that of protons implies that volume effects (and/or projectile
fragmentation) are significant in deuteron production.

{\it Tritons, $^3$He, and $^4$He}.
The results on anisotropic flow of light nuclei in the beam rapidity
region are presented in Fig.~4. 
Contamination between different species in this data sample is less than 15\%.
Directed flow monotonically increases with particle mass and reaches
values up to about 0.8 for high $p_t$ particles. 

\vspace{-2mm}
\centerline{
\psfig{figure=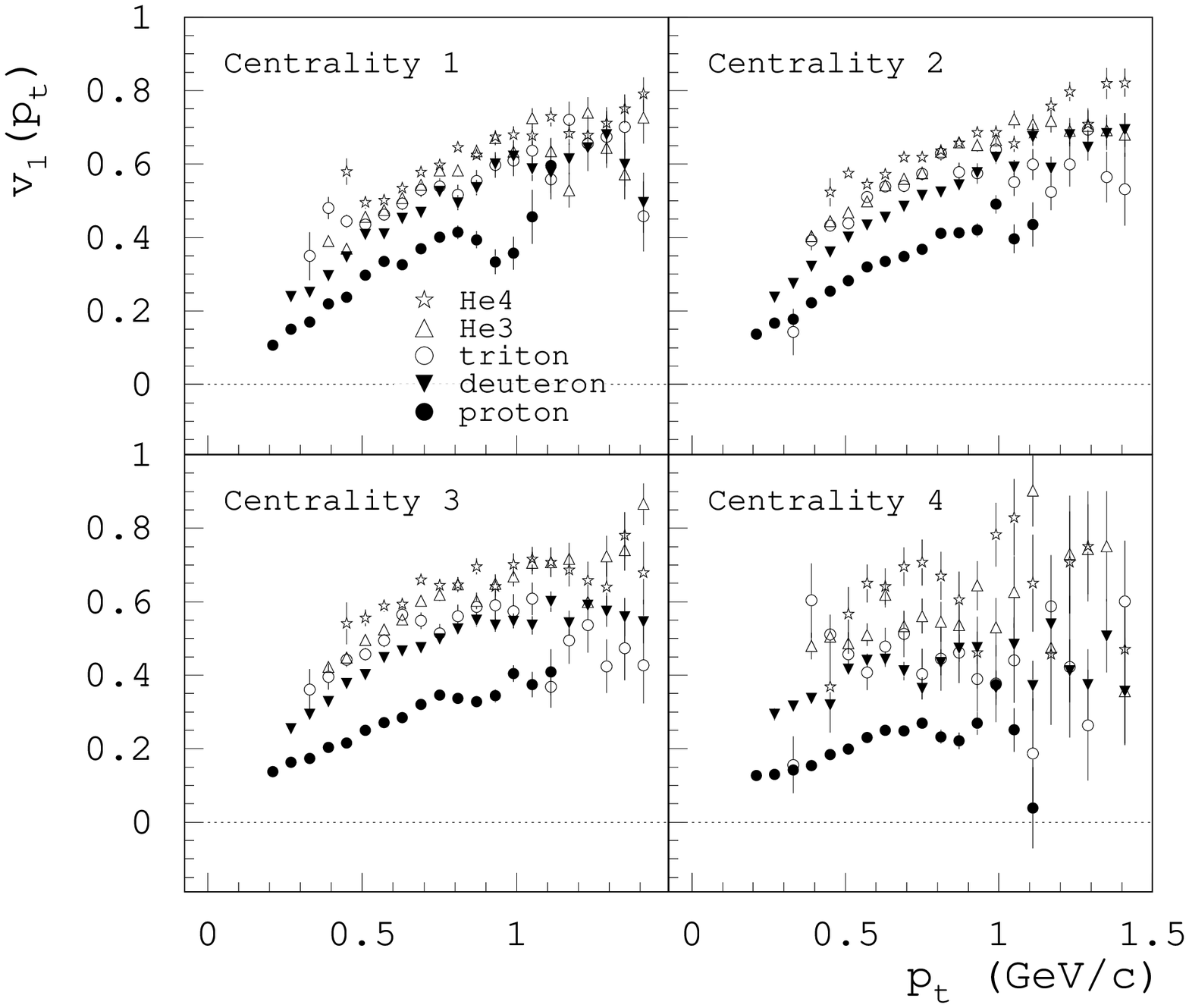,height=12.5cm}
}
\vspace{-11mm}
\centerline{{\small Figure 4.  Directed flow of light nuclei in the rapidity
region $3.0<y<3.2$}}
\smallskip

\section{Summary}

The transverse momentum dependence of directed flow has been studied
in Au+Au collisions at the BNL AGS for different types of particles:
protons, charged pions, $K^+$, and light nuclei.
Even a brief look at these data reveals rich and interesting physics.
At present, the quality of the experimental data in this field is 
more advanced than theory and good/detailed models are needed for 
the interpretation of the data.

\medskip

This research was supported, in part, by the U.S. DOE, the NSF, 
and Natural Sciences and Engineering Research Council of Canada.

\end{document}